\documentclass[10pt,twocolumn,a4paper]{article}
\usepackage[T1]{fontenc}
\usepackage[cp1250]{inputenc}
\usepackage{graphicx}
\usepackage{epstopdf}

\usepackage{varioref}
\usepackage{amsmath,amssymb,amsfonts}

\usepackage[pdftex,breaklinks=true]{hyperref}
\hypersetup{
colorlinks=true,
linkcolor=black,          
citecolor=black,        
filecolor=black,      
urlcolor=black,           
bookmarks=true,
bookmarksnumbered=true,
pdfauthor={Piotr Bania},
pdfstartview=FitH,
pdfsubject={Evading network-level emulation},
pdftitle={Evading network-level emulation}
}

\usepackage{breakurl}
\usepackage{url}
\usepackage{algorithmic}
\usepackage{listings} 
\usepackage{color} 
\usepackage[ruled,vlined]{algorithm2e}

\title{Evading network-level emulation}
\author{Piotr Bania\\
\texttt{\href{mailto:bania.piotr@gmail.com}{bania.piotr@gmail.com}}}
\date{April 2009}

\begin{document}
\maketitle

\begin{abstract}
Recently more and more attention has been paid to the intrusion detection systems (IDS) which don't rely on signature based detection approach. Such solutions try to increase their defense level by using heuristics detection methods like network-level emulation. This technique allows the intrusion detection systems to stop unknown threats, which normally couldn't be stopped by standard signature detection techniques. 

{\noindent\newline}In this article author will describe general concepts of network-level emulation technique including its advantages and disadvantages (weak sides) together with providing potential countermeasures against this type of detection method.  

\end{abstract}

\section{Introduction}

Intrusion detection systems were designed to detect and deny unauthorized access attempts launched mainly through a network. Together with the growth of the Internet number of such attempts increased dramatically. Most of the known network attacks are designed directly to compromise the security of a targeted computer system. This includes variety of hacking attempts against vulnerable services, unauthorized access to sensitive data and all types of malware (viruses, worms etc.). We have already seen worms that infected millions of computer systems in a very short period of time, like Blaster worm which infected more the 25 million unique computer machines \cite{blaster_article}. Intrusion detection systems objective is to detect such attacks and take all the necessary actions to prevent further spread. 

{\noindent\newline}Most of the IDS still rely on the signature (pattern) matching approach. Basically they are constantly listening to the network traffic and trying to find specific signature inside the packet data. When a potentially malicious pattern is found, intrusion detection system blocks the packet from going "deeper" into the network. The signature detection approach often plagues IDS with high number of false positives alerts \cite{ids_detection_datamining,ids_false_positive}. In fact, there are many IDS attacks which make a nasty usage of this issue (for example the old squealing attack \cite{Patton01anachilles}). \\
{\noindent\newline}The pattern matching approach was (and in fact still is) heavily used in antivirus software. Somewhere in the end of 80s (early 90s) first polymorphic viruses started to appear \cite{virus_history1, virus_history2}. The 1260 virus (also known as Chameleon) is considered as the first polymorphic virus. The 1260 virus, created by Mark Washburn, was in fact a fusion between the Vienna virus (written, and then published by Ralf Burger in his book) and the Cascade virus (the first self-encrypting file virus). Washburn extended the initial Cascade virus technique, which resulted in the creation of decryptor with mutable body. Decryptor body was generated (changed) upon infection - of course the general logic of decryption algorithm was preserved. Soon after this more and more complex polymorphic engines started to appear, like the Mutation Engine (MtE) (created by Dark Avenger) which appeared in 1992 or the DAME engine (created by Dark Angel from Phalcon/Skism group), which appeared in 1993 as a part of 40Hex magazine. Summing it up, polymorphism is a technique that allows to evade signature detection by encrypting code and creating a decryptor (decrypting stub) which is different every generation\footnotemark. {\footnotetext{Author is refering here to the so called the fast-polymorphism, not the slow-polymorphism approach where the main goal is to limit the number of mutated samples - this complicates the process of signature creation.}} Polymorphism was applied multiple times \cite{NIDS_poly_evasion,CLET,ADmutate,Tapion} to shellcode generation process. Polymorphic shellcodes were and still are the ultimate weapon against intrusion detection systems based only on signature detection approach. However, like in the antivirus software case, some countermeasures were developed in order to stop polymorphic code. For this purpose code emulation technique was presented. Code emulation is a powerful technique used already for a long time in the antivirus software field (especially in dealing with polymorphic viruses with encrypted bodies). Considering the fact the decrypting procedure (decryptor) must decrypt the actual body of the virus before it actually get executed, an code emulator may be used to simulate the work of decryptor. When the original virus body is decrypted antivirus software may proceed with standard pattern matching detection. Of course this is the most optimistic scenario. Not so far ago a group of researchers \cite{ids_emu1,ids_emu2} proposed similar solution for the purpose of intrusion detection systems. In their approach called network-level emulation they suggested using a CPU emulator to dynamically analyze every potential instruction sequence in the monitored network traffic in order to locate execution behavior of potentially malicious code. The network-level emulation does not rely on pattern matching approach, what allows it to work like an heuristics scanner and detect previously unknown threats. 

{\noindent\newline}In this article, the author will describe general assumptions used by intrusion detection systems in terms of detecting attacks via using network-level emulation technique. Potential evasion techniques will be presented as well. To the author's knowledge the only emulation based "IDS" publicly available is a library created by Paul Baecher and Markus Koetter called libemu \cite{libemu}. Libemu library is developed more like a proof of concept than actual stable working product, so it's functionality is quite limited. Due to the lack of others real testing environments, anti-emulation techniques presented in this article should be treated mostly as theoretical concepts. Bypassing techniques presented in this paper will focus only (mostly) on defeating network-level emulation approach. This paper will also focus on Windows shellcodes, because they are naturally bigger and more complicated then ones designed to work in *nix systems. Author will also refer to shellcode as a type of shellcode which requires a decryptor.

\section{Concepts of network-level emulation}
In order to perform code emulation intrusion detection system necessary data must be gathered first. To achieve this IDS monitors client-initiated data of the network stream. Typically such data may include malicious requests like exploitation attempts. For TCP packets the application-level stream is reassembled when necessary and in case of large client-initiated streams only some limited portion of it is inspected. When the information is collected the emulation process begins. Since the shellcode position is not known at this point and the lengths of IA-32 instructions may differ, emulator must treat every byte in the data buffer as a potential entrypoint of the shellcode. Basically the emulation process is repeated for each byte position found in the gathered data buffer. Depending on the emulated code behavior it is marked as potential shellcode or not. The detection method will be discussed in the next section.  

\subsection{Shellcode Detection}
This section will present general shellcode architecture and also the heuristics shellcode detection mechanisms used in network-level emulation. 

\subsubsection{Shellcode Architecture}
\label{sec:shell_arch}
\begin{figure}[tbhp]
\centering
\includegraphics[scale=1]{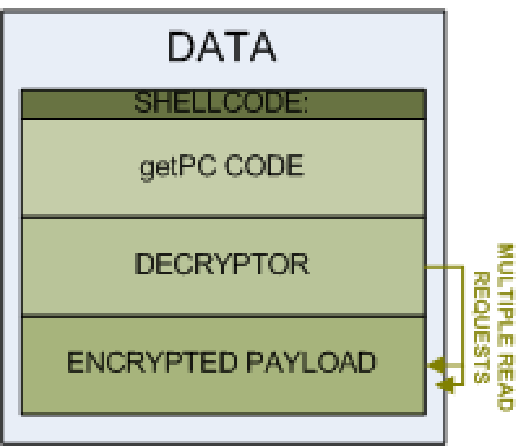}
\caption{General composition of typical Windows shellcode.}
\label{img:shellcode_parts}
\end{figure}

Typical Windows shellcode consists of three parts: GetPC code, decryptor (decrypting stub), encrypted payload (\autoref{img:shellcode_parts}).\newline

{\noindent}Where:
\begin{itemize}
    \item {\textbf{GetPC code}} -  this part is important because shellcode decryptor must calculate the correct address of the encrypted payload - otherwise the decryption process will be done wrong. Since the final location of injected shellcode cannot be predicted and on the other hand IA-32 architecture does not provide {\texttt{EIP}}-relative addressing mode the shellcode address must be calculated on-the-fly (while the shellcode executes)\footnotemark. \footnotetext{Same mechanisms are also used in computer viruses.}
 
    \item {\textbf{decryptor}} - this part is responsible for decrypting the encrypted payload. In order to decrypt the payload, decryptor must know its location (this is provided by GetPC code), must be able to read the data from the payload and also must be able to write the decrypted code back (usually to the same place). A typical decryption algorithm is a implementation of simple XOR cipher. 

    \item {\textbf{encrypted payload}} - this is the effective part of the shellcode. It performs all the necessary actions to satisfy the attacker, like binding a shell to a specific port etc. In most of the cases the payload must be encrypted to avoid using restricted byte characters like NULL-bytes - this is limited by the vulnerability itself. Once the original payload data is decoded, the decoder stub passes execution to it.   
\end{itemize}

\subsubsection{Detection Methods (heuristics)}
\label{sec:detection}
As it was presented in the previous section, encrypted shellcode payload must be decrypted in order to retrieve its original functionality. The decryption process itself depends on the body of encrypted payload - which is obvious. Therefore the decryptor always must make read requests to the body of encrypted payload. This is the main assumption of the network-level emulation technique \cite{ids_emu1,ids_emu2}. Additionally, the read requests caused by the decryptor must refer to the memory located in range of current data buffer (basically to the contents of shellcode). This was intially the only assumption in this detection method. However it appears that sometimes random code can cause hundreds of read requests of different memory locations in range of the data buffer. This may lead to high number of false positive alerts. Since this assumption alone was not strict enough, additional one was created. As it was stated before (section \ref{sec:shell_arch}), typical shellcode uses GetPC code in order to obtain its virtual address in the memory. This assumption is used together with the previous one (decryptor needs to make read requests). Whenever in the execution path a GetPC code block is found and it is followed by read requests to a memory location in range of the data buffer, the data buffer is marked as shellcode. Appending to the tests \cite{ids_emu1}, by basing the detection method on those assumptions the network-level emulator was able to detect all common polymorphic shellcodes without making false positives alerts.

\section{Limitations and countermeasures}
Even though the network-level emulation technique is able to detect most of the currently known polymorphic shellcodes generated by various tools like ADmutate \cite{ADmutate}, CLET \cite{CLET}, TaPiON \cite{Tapion} there are still some serious issues that should be taken into consideration. This section will list some of them. Some of the issues presented here have already been mentioned \cite{ids_emu1}, however for the sake of understanding the general network-level emulation limitations they are presented and described more deeply in this section as well. 

\subsection{Address Space and CPU Context State Problem}
\label{addr_and_cpu_problem}
The main problem of the emulation approach is that it cannot provide 100\% correct memory view of the process address space. Furthermore, it seems to be impossible to find a solution for this issue, especially in case of intrusion detection systems installed on the separate machines. Of course one of the potential short-term solutions would be to statically map some of the mostly used Windows libraries like {\texttt{KERNEL32.DLL}} or {\texttt{NTDLL.DLL}}. However this is a very limited solution and not really satisfactory. It is highly probable that more skilled attacker would know his target well, so in case of exploiting an specific process he may not reveal the decryption routine code if for example some of the additional libraries (typically used by the targeted process) are missing (this can be easily achieved by parsing PEB (Process Enviroment Block) \cite{RatterPEB,skape_shellcode,LSD_shellcode}). It is very unlikely that any intrusion detection system based on the network-level emulation approach will be able to solve such issues, since typically such systems are completely separated from the target process. Same goes for the state of CPU context. Typically attacker can assume some of the registers values - even if not accurately the approximate value still can be calculated. Since emulators are separated from the running processes, they are unable to guess correct values. For example libemu \cite{libemu} nullifies all registers values (except {\texttt{ESP}}) before starting the emulation process, and some other solution \cite{ids_emu1} keeps all of the registers values randomized. Both of presented methods can be defeated very easily, since it is very unlikely that generated random values will meet ones from the attacked process.

\subsection{GetPC Code}
The GetPC code is often very important to shellcode. In fact most of the known shellcodes totally rely on it. The GetPC code is usually implemented in two ways, by using relative {\texttt{CALL}} instruction or by using the {\texttt{FSTENV}} instruction. The implementation used in libemu \cite{libemu} scans for either {\texttt{CALL rel}} ({\texttt{0xE8}}) opcode and checks if the call destination resides in data buffer or scans for the {\texttt{FSTENV}} instruction. If even one of the variants is located then code is marked as potentially harmful. The libemu implementation is not really accurate since instead of  {\texttt{CALL rel}} instruction, a  {\texttt{CALL indirect}} may be used (where the operand can be a memory location or a register). For example following sequence of instructions (see Listing \ref{delta1}) would initialize {\texttt{ESI}} register with the address pointing to the first instruction after {\texttt{CALL}}.\newline 

{\ttfamily{\footnotesize{
\lstset{language={[x86masm]Assembler}}
\begin{lstlisting}[frame=trbl, label=delta1, caption={Variation of GetPC code using indirect {\texttt{CALL}} instruction.}, captionpos=b]{}
PUSH 	0C390565Eh
CALL	ESP
\end{lstlisting}
}}}

{\noindent}Other methods may be used as well, like the SEH (Structured Exception Handler) method\footnotemark \cite{getPC}. Furthermore, in some cases the GetPC code can be simply omitted, by assuming one of the registers or any other element from the attacked process address space, point to the shellcode base address (or somewhere near). Since emulator cannot predict such values correctly (see section \ref{addr_and_cpu_problem} for details) this is a very good and nasty evasion technique. \footnotetext{Matter of fact this technique is often used for anti-debugging, anti-emulator purposes especially by executable file packers.} For example if attacker can assume his shellcode will be located somewhere at the stack space of targeted module, he can use following algorithm to find the shellcode memory address:
\begin{enumerate}
    \item grab top stack address ({\texttt{FS:[0x04]}}) and bottom stack address ({\texttt{FS:[0x08]}}) from TIB (Thread Information Block)
    \item scan the obtained stack memory range for a shellcode marker
\end{enumerate}

{\noindent}Presented method does not rely on hard coded values and practically can be applied to any Windows shellcode (of course shellcode must be located at the stack). Important fact is even if the shellcode scans the stack memory space (makes a read requests) it still does not trigger the IDS alert, because it does not fulfill the assumptions presented in section \ref{sec:detection}.

\subsection{Read requests}
As it was previously presented (section \ref{sec:detection}) when GetPC code is missing in the gathered data buffer there is no need for the decryptor to hide the read requests. However even with detected GetPC code it is still possible to bypass the read requests detection technique. This can be achieved by moving the original shellcode data to other location (aside from data buffer) like to an allocated memory space. Problem of this technique is that in order to copy the shellcode to other location decryptor must still be able to read it - unfortunately such action will result in setting of the IDS alarm. The trick here is to force other system components to make the read access instead of doing this directly from the shellcode decryptor. This can be achieved by variety of tricks since a lot of {\texttt{memcpy}} alike functions are found in most of the Windows libraries. The code presented below (Listing \ref{antiread}) uses native API \cite{native_api,syscall_shellcode} to allocate memory and copy itself to the allocated region. In this example {\texttt{ZwAllocateVirtualMemory}} (syscall \#11h) function is used to allocate necessary memory space and {\texttt{ZwReadVirtualMemory}} (syscall \#BAh) is used to copy the shellcode to the newly obtained location. Both syscall numbers were targeted for the Microsoft Windows XP SP2 operating system.\newline

{\ttfamily{\footnotesize{
\lstset{language={[x86masm]Assembler}}
\begin{lstlisting}[frame=trbl, label=antiread, caption={Assembly pseudo-code which can be used to evade payload-read detection.}, captionpos=b]{}
xor	eax,eax
push	PAGE_READWRITE_EXECUTE
push	MEM_COMMIT
push	offset region_size_ptr
push	eax
push	offset out_base
push	-1
push	offset ret1
push	offset ret1
mov	edx,esp
mov	eax,011h
sysenter
ret1:

push	eax
push	REGION_SIZE
push	dword ptr [out_base]
push	offset shellcode
push	-1
push	offset ret2
push	offset ret2
mov	edx,esp
mov	eax,0BAh	
sysenter
ret2:
\end{lstlisting}
}}}

{\noindent}After the shellcode relocation execution can be continued from the newly allocated memory or just the decryptor may refer to the allocated memory while performing the decryption instead of using the code from the data buffer. It's obvious that in this example if the native API functions will not be simulated correctly future decryption process will fail. 

\subsection{Time Limit}
An emulator, no matter if it is used in the intrusion detection systems or antivirus software products, must do the analysis in fixed period of time. This is crucial and not possible to avoid. Generally code emulation is hundreds times slower then native execution. This is one of the biggest disadvantages of the emulation approach. In the world of computer viruses several anti-emulation techniques were deployed to make use of this issue, like for example the branching technique \cite{MentalBranching}. However shellcodes unlike viruses are very limited in size so in this case techniques like branching (which require a lot of additional space) are not very useful. Some of the known polymorphic shellcode generators \cite{Tapion} are using the {\texttt{RDTSC}} (Read Time Stamp Counter) instruction to detect debuggers or some of the emulators. However emulator may simulate the {\texttt{RDTSC}} instruction and stay undetected. The most basic and stable attempt in this case would be to place several delaying loops which would iterate enough times until the finite time given for emulator to work will not be consumed. Appending to the report \cite{ids_emu1} additional heuristics were applied to detect endless (infinite) loops which are sometimes found in the random code. However proposed heuristic methods can be easily defeated and basically a specially crafted endless loop may be used to consume entire emulator time.

\subsection{Other Techniques}
There are quite a lot of other techniques that can be used for bypassing network-level emulators. For example, even if emulator traces all the memory writes to the data buffer it can only predict situations where the memory is not written from external components (like for example the operating system kernel itself). So if shellcode would use an native API function like {\texttt{ZwWriteVirtualMemory}} or any other suitable function to perform memory write, emulator would not be able to correctly reproduce the changes. Using the Prefetch Input Queue (PIQ) \cite{prefetchinputque} tricks often cause a corruption of the emulated code, because tis implementation is not done correctly (like self-overwriting {\texttt{REP STOSB}} etc.). Additionally it appears none of the network-level emulators can properly emulate FPU or SSE instructions, so in case where the shellcode depends\footnotemark\ on the results given by FPU, SSE instructions it highly probable that emulation process will fail.\footnotetext{TaPiON \cite{Tapion} generates a decryptor together with FPU instructions as well, however they are not a part of the decrypting process - they are used as {\texttt{NOP}} instructions. This is not a good solution because emulator may just ignore such instructions and still the decryption process will succeed.} The swarm attack \cite{Swarm} is also used against network-level emulation systems. The main idea of the swarm attack is to modify a control hijacking attack so that the shellcode decoder will not be present in any attack traffic. This is achieved by building the decoder inside the attacked process from a small pieces which are send by multiple attack instances.   
 
\section{Acknowledges}
Author would like to thank Julien Vanegue and Małgorzata G. for helping with writing this article.

\section{Conclusion}
In this paper the author presented main concepts of network-emulation technique, together with its advantages and weak spots. Appending to the reports \cite{ids_emu2} real-world network-level emulation implementation provided good overall results in detecting known polymorphic shellcodes. This paper illustrated potential evasion techniques which would make the detection harder or just impossible. It's obvious that together with the growth of instruction detection systems based on network emulation approach presented evasion techniques will be applied to shellcodes creation process as well. From the other hand network-level emulation solution requires a lot of processing power so it is very unlikely it will relieve the signature detection based solutions soon if ever. Anyway time it occurs we will be prepared. 

\newpage
\bibliographystyle{plain}
\bibliography{F:/latex_sheets/spiderpig_thesis/spiderpig/bibliografia}

\begin{thebibliography}{10}

\bibitem{virus_history1}
{Alberto Escudero Pascual}.
\newblock {History of Computer Viruses}.

\bibitem{libemu}
Paul Baecher and Markus Koetter.
\newblock {Libemu - shellcode detection library}.
\newblock \url{http://libemu.carnivore.it}.

\bibitem{Tapion}
Piotr Bania.
\newblock {TaPiON} - {Polymorphic Deciphering Algorithm Generator Project}.
\newblock \url{http://www.piotrbania.com/all/tapion/}.

\bibitem{syscall_shellcode}
Piotr Bania.
\newblock {Windows Syscall Shellcode}.
\newblock \url{http://www.securityfocus.com/infocus/1844/1}.

\bibitem{Swarm}
Simon~P. Chung and Aloysius~K. Mok.
\newblock {Swarm Attacks against Network-Level Emulation/Analysis}.
\newblock In {\em {RAID '08: Proceedings of the 11th international symposium on
  Recent Advances in Intrusion Detection}}, pages 175--190, Berlin, Heidelberg,
  2008. Springer-Verlag.

\bibitem{MentalBranching}
The~Mental Driller.
\newblock Advanced polymorphic engine construction.
\newblock {\em 29A Magazine Issue \#5}, 2000.

\bibitem{NIDS_poly_evasion}
Yuri Gushin.
\newblock {NIDS polymorphic evasion - The End?}
\newblock \url{http://www.milw0rm.com/papers/18}.

\bibitem{getPC}
Costin Ionescu.
\newblock {GetPC} code.
\newblock \url{http://www.securityfocus.com/archive/82/327348/2009-04-22/1}.

\bibitem{virus_history2}
{Joe Wells}.
\newblock {Virus Timeline}.
\newblock \url{http://www.research.ibm.com/antivirus/timeline.htm}.

\bibitem{ADmutate}
K2.
\newblock {ADMmutate}, a shellcode mutation engine.
\newblock \url{http://www.ktwo.ca/security.html}.

\bibitem{blaster_article}
Robert Lemos.
\newblock {MSBlast infected more than 25 million}.
\newblock \url{http://www.securityfocus.com/brief/72}.

\bibitem{native_api}
Gary Nebbett.
\newblock {\em {Windows NT/2000 Native API Reference}}.
\newblock Sams - PEARSON, 2000.

\bibitem{LSD_shellcode}
Last~Stage of~Delirium.
\newblock Win32 assembly components.

\bibitem{Patton01anachilles}
Samuel Patton, William Yurcik, and David Doss.
\newblock {An Achilles' Heel in Signature-Based IDS: Squealing False Positives
  in SNORT}.
\newblock
  \url{http://www.scs.carleton.ca/~soma/id-2007w/readings/patton_yurcik_doss_r%
aid2001.pdf}, 2001.

\bibitem{ids_detection_datamining}
Tadeusz Pietraszek and Axel Tanner.
\newblock {Data Mining and Machine Learning---Towards Reducing False
  Positives}.
\newblock \url{http://tadek.pietraszek.org/publications/pietraszek05_data.pdf}.

\bibitem{ids_emu1}
Michalis Polychronakis, Kostas~G. Anagnostakis, and Evangelos~P. Markatos.
\newblock Network-level polymorphic shellcode detection using emulation.
\newblock In {\em In Proceedings of the GI/IEEE SIG SIDAR Conference on
  Detection of Intrusions and Malware and Vulnerability Assessment (DIMVA)},
  pages 54--73, 2006.

\bibitem{ids_emu2}
Michalis Polychronakis, Kostas~G. Anagnostakis, and Evangelos~P. Markatos.
\newblock Real-world polymorphic attack detection using network-level
  emulation.
\newblock In {\em CSIIRW '08: Proceedings of the 4th annual workshop on Cyber
  security and information intelligence research}, pages 1--3, New York, NY,
  USA, 2008. ACM.

\bibitem{RatterPEB}
Ratter.
\newblock Gaining important datas from {PEB} under {NT} boxes.
\newblock {\em 29A Magazine Issue \#6}, 2001.

\bibitem{skape_shellcode}
skape.
\newblock {Understanding Windows Shellcode}.
\newblock \url{http://www.hick.org/code/skape/papers/win32-shellcode.pdf}.

\bibitem{CLET}
CLET Team.
\newblock Polymorphic shellcode engine using spectrum analysis.
\newblock {\em Phrack Magazine Issue \#65}, 2003.

\bibitem{prefetchinputque}
Wikipedia.
\newblock {Prefetch input queue}.
\newblock \url{http://en.wikipedia.org/wiki/Prefetch_input_queue}.

\bibitem{ids_false_positive}
Emmanuele Zambon and Damiano Bolzoni.
\newblock {Network Intrusion Detection Systems. False Positive Reduction
  Through Anomaly Detection}.
\newblock
  \url{http://www.blackhat.com/presentations/bh-usa-06/BH-US-06-Zambon.pdf}.

\end{thebibliography}
\end{document}